\documentclass[twocolumn,showpacs,aps,prl,superscriptaddress,floatfix]{revtex4}
\usepackage{graphicx}
\usepackage{dcolumn}
\usepackage{bm}

\begin{document}

\preprint{}

\newcommand{\tto}{Tb$_{2}$Ti$_{2}$O$_{7}$ }
\newcommand{\yto}{Y$_{2}$Ti$_{2}$O$_{7}$ }
\author{J.P.C. Ruff}
\affiliation{Department of Physics and Astronomy, McMaster University,
Hamilton, Ontario, L8S 4M1, Canada}
\author{B.D. Gaulin}
\affiliation{Department of Physics and Astronomy, McMaster University,
Hamilton, Ontario, L8S 4M1, Canada} 
\affiliation{Canadian Institute for Advanced Research, 180 Dundas St. W.,
Toronto, Ontario, M5G 1Z8, Canada}
\author{J.P. Castellan}
\affiliation{Department of Physics and Astronomy, McMaster University,
Hamilton, Ontario, L8S 4M1, Canada}
\author{K.C. Rule}
\affiliation{Department of Physics and Astronomy, McMaster University,
Hamilton, Ontario, L8S 4M1, Canada}
\author{J.P. Clancy} 
\affiliation{Department of Physics and Astronomy, McMaster University, 
Hamilton, Ontario, L8S 4M1, Canada}
\author{J. Rodriguez}
\affiliation{Department of Physics and Astronomy, McMaster University,
Hamilton, Ontario, L8S 4M1, Canada}
\author{H.A. Dabkowska} 
\affiliation{Department of Physics and Astronomy, McMaster University, 
Hamilton, Ontario, L8S 4M1, Canada}

\title{Structural Fluctuations in the Spin Liquid State of \tto } 

\begin{abstract} 

High resolution X-ray scattering
measurements on single crystal {\tto} reveal finite structural
correlations at low temperatures.  This geometrically frustrated
pyrochlore is known to exhibit a spin liquid, or cooperative paramagnetic
state, at temperatures below $\sim$ 20 K.  Parametric studies of
structural Bragg peaks appropriate to the Fd$\bar{3}$m space group of 
\tto
reveal substantial broadening and peak intensity reduction in the
temperature regime 20 K to 300 mK.  We also observe a small, anomalous
lattice expansion on cooling below a density maximum at ${\sim}$ 18 K.  
These measurements are consistent with the development of fluctuations
above a cooperative Jahn-Teller, cubic-tetragonal phase transition at very
low temperatures.

\end{abstract} 

\maketitle 


A magnetic material whose lattice geometry
prevents the simultaneous satisfaction of the local magnetic interactions
is said to be geometrically frustrated.  Such materials, many of which are
based on triangular and tetrahedral crystalline architectures, are of
intense current interest because they have a natural proclivity towards
exotic quantum mechanical ground states \cite{Ramirez94}.  As the leading
magnetic interactions on such lattices are frustrated, rather subtle
subleading terms in the Hamiltonian, such as interactions beyond
nearest-neighbours, weak disorder, and the order-by-disorder mechanism,
tend to determine the ultimate ground state of the material.  

The rare-earth titanates, with formula A$_2$Ti$_2$O$_7$, are a 
well-studied
family of frustrated magnetic insulators.  In this family, the A-site is
occupied by a trivalent rare-earth ion with eight-fold oxygen
coordination, while the Ti$^{4+}$ ion has six-fold oxygen coordination.  
Both the A$^{3+}$ site and the Ti$^{4+}$ site independently reside on the
pyrochlore structure, a face-centered cubic lattice of corner-sharing
tetrahedra which is a playground for phenomena related to geometrical
frustration.  Depending on the nature of the magnetic rare-earth ion in
these materials, their ground states can exhibit long-range magnetic order
\cite{ChampionPRB, RajuGTO}, spin ice physics \cite{Harris97,denHertog00},
and in the case of {\tto}, a highly correlated quantum disordered state
known as a collective paramagnet or spin liquid
\cite{Gardner99,Gardner01,Gardner03}.  

The low temperature properties of \tto
have been extensively studied. However, the crystal lattice is
generally assumed to be a passive bystander to the spin liquid physics.  
In this letter, we present high resolution x-ray scattering evidence for
strong fluctuations in the low temperature lattice of {\tto}, signifying
both a feedback of frustration onto the crystal lattice, and the
importance of the lattice degrees of freedom in determining the exotic
ground state of this material.

\begin{figure} 
\centering 
\includegraphics[width=8.5cm]{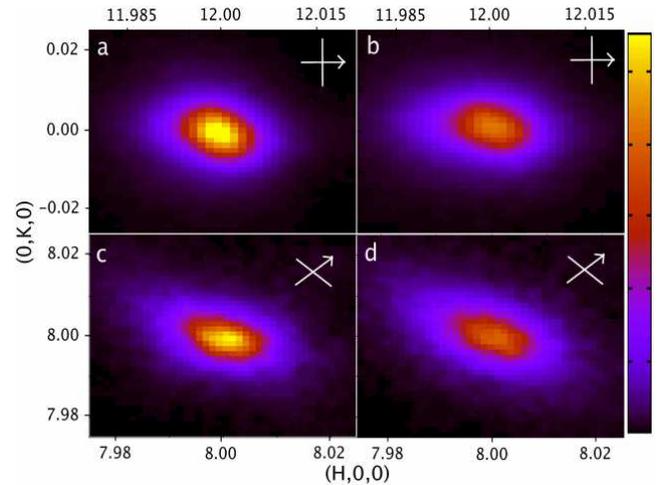}
\caption{(H,K,0) reciprocal space maps of structural Bragg peaks of \tto.
(a) (12, 0, 0) at T=40 K. (b) (12, 0, 0) at T=0.3 K. (c) (8, 8, 0) at T=40 
K.  
(d) (8, 8, 0) at T=0.3 K.  Logitudinal (arrowhead) and transverse 
directions
in reciprocal space are indicated in each panel. The low temperature
broadening of the Bragg peaks is preferentially longitudinal for (12, 0,
0) and preferentially transverse for (8, 8, 0). } 
\label{Figure 1}
\end{figure}

Current interest in {\tto} is largely due its failure to attain magnetic
order of any kind to temperatures as low as 20mK, despite having an
antiferromagnetic (AFM) Curie-Weiss constant of $\sim$ 18 K
\cite{Gardner99,Gardner01,Gardner03}.  Magnetic neutron scattering on \tto
at low temperatures\cite{Gardner03} shows diffuse scattering indicative of spin correlations over the spatial
extent of a single tetrahedron ($\sim$ 3.6 \AA) in zero magnetic field and
ambient pressure.  When perturbed with either relatively weak magnetic
fields or applied pressures, or both, long range magnetic order can be
induced at temperatures of $\sim$ 1-3 K
\cite{rule06,Mirebeau05,MirebeauPF}.  

Low lying crystal field levels for Tb$^{3+}$ in this environment have been
measured\cite{GingrasCF} and they are consistent with local [111]
anisotopy, such that Tb moments are constrained to point into or out of
the tetrahedra.  However, the diffuse magnetic neutron
scattering\cite{Gardner01} as well as the presence of dispersive spin
waves in the magnetic field-induced ordered state\cite{rule06} of \tto at
low temperatures are more easily understood in terms of continuous spin
symmetries.  This quandry has been been recognized and discussed for some
time\cite{Enjalran04,Molavian06}.

Much earlier magneto-elastic measurements on \tto revealed both giant
magnetostriction\cite{Aleksandrov} and an anomolous Young's modulus
\cite{Mamsurova} at low temperatures, indicating the importance of the
coupling between magnetic and lattice degrees of freedom.  Strong
magnetoelastic effects have also been topical in cubic spinel
antiferromagnets. The spinels have chemical composition AB$_2$O$_4$, and
the B-sublattice ions also reside on a network of corner-sharing
tetrahedra; that is the pyrochlore lattice. Strongly first order
structural and magnetic phase transitions can occur simultaneously in some
of these materials at low temperatures.  In the case of ZnCr$_2$O$_4$ this
occurs near 12.5 K, wherein the low temperature lattice undergoes a
tetragonal distortion \cite{Lee00, Lee02}.  Diffuse quasi-elastic magnetic
scattering due to dynamic, short range spin correlations characteristic of
geometrical frustration is associated with the high temperature cubic
phase.  This quasi-elastic magnetic scattering becomes gapped within the
low temperature tetragonal phase of ZnCr$_2$O$_4$, drawing similarities to
spin-Peierls physics, albeit in three dimensional materials.  Other
spinels exhibit unexpected plateaus in their magnetization as a function 
of field at low
temperature which have been of intense interest
\cite{Ueda05,Matsuda07}.  Such magnetization plateaus have been modeled
with biquadratic exchange terms in the effective spin Hamiltonian of these
materials, arising from strong magneto-elastic coupling \cite{Penc07}.
Theoretically, spin-Peierls-like distortions of the pyrochlore lattice
which act to better satisfy frustrated AFM interactions have been
predicted \cite{Tchern02}.  No such transition has been observed in {\tto}
despite several systematic studies \cite{Ofer07,Han04}.



Large single crystals of {\tto} and {\yto} were grown at McMaster
University using the traveling solvent floating zone technique, and a
two-mirror, NEC image furnace.  Both crystals were grown at a rate of 6 
mm/h in an Ar environment.  The growth
protocol was similar to that originally reported for {\tto}
\cite{Gardner98}.  Samples with aproximately cubic dimensions and 1 mm$^3$
volumes were subsequently cut from the larger growths for use in the x-ray
measurements.  The samples were aligned with their (H, K, 0) plane
coincident with the scattering plane, and placed in a Janis $^3$He
cryostat mounted within a four-circle x-ray diffractometer.  Cu
K${\alpha}_1$ radiation from an 18kW rotating anode x-ray generator was
selected using a perfect single crystal germanium (110) monochromator.  
The diffracted beam was measured using a Bruker Histar area detector, 
mounted on the scattering arm at a distance
of 0.7 m from the sample.  This configuration allows for high-resolution
characterization of Bragg peaks from room temperature down to ${\sim}$ 300
mK, with a temperature stability of ${\sim}$ 10 mK.



\begin{figure}
\centering
\includegraphics[width=8.5cm]{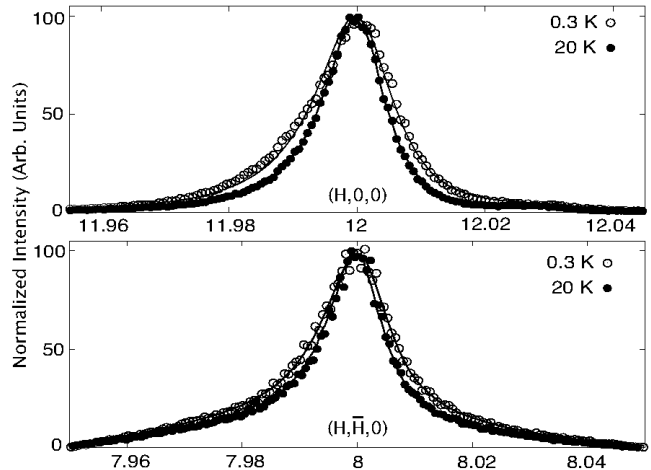}
\caption{Cuts through the Bragg peaks, at both high and low temperatures, 
along with fits (solid lines) of the data to resolution convoluted 
Lorentzians, as described in the text.  Intensities have been normalized to 
highlight the change in width.  Top:  Longitudinal broadening of the (12, 
0, 0) 
peak.  Bottom:  Transverse broadening of the (8, 8, 0) peak.}
\label{Figure 2}
\end{figure}

Measurements were carried out on both {\tto} and {\yto} as a function of
temperature for two high-angle Bragg reflections in the (H, K, 0) plane, 
the
(12, 0, 0) and (8, 8, 0) Bragg peaks.  Representative two-dimensional maps
at T=40 K and T=0.3 K are shown in Fig. 1.  The longitudinal and
transverse (relative to the relevant reciprocal lattice vector) directions
in reciprocal space are indicated by the arrows and straight 
lines, respectively, superposed on the figure.  
At low temperatures, the peak intensities of both reflections
are suppressed and the width of the scattering substantially increases as
compared to the same data at 40 K.  A careful inspection of
the data shows that the (12, 0, 0) Bragg peak broadens preferentially in 
the
longitudinal (H, 0, 0) direction, while the (8, 8, 0) reflection broadens
preferentially in the transverse (H, -H, 0) direction.  In each case, the
integrated intensity of the peaks in Fig. 1 shows no temperature
dependence.  

Cuts through these maps at T=20 K and T=0.3 K are shown
in Fig. 2, highlighting the dominant broadening for each peak.  
In both cases, the peak intensities have been normalized and it is clear
that the peak widths increase substantially as the temperature is lowered
into the spin liquid regime appropriate to \tto.  Similar high resolution x-ray scattering measurements were carried out on the
non-magnetic pyrochlore \yto, which shows no corresponding anomalous
broadening or Bragg intensity variation.  We conclude that \tto develops
finite spatial lattice correlations below $\sim$ 20 K. We will argue that
this is consistent with a system approaching a cooperative Jahn Teller
phase transition - albeit at unobtainably low temperatures.

The top panel of Fig. 3 shows the temperature dependence of the peak intensity at the (12,
0, 0) and (8, 8, 0) Bragg positions in \tto. The temperature dependence at
the (12, 0, 0) Bragg position in \yto is also shown for comparison.  Both
Bragg intensities in \tto decrease by about 20\% between 20 K and 0.3 K,
and the decrease is continuous.  No corresponding effect is seen in \yto.  
The data shown in Figs. 1 and 2 was also fit to an Ornstein-Zernike form
\cite{Collins} for spatial correlations (Eq. 1) in order to extract
inverse correlation lengths along both the longitudinal and transverse
directions.

\begin{equation}
S(q) = \frac{A}{1 + (\frac{q-q_o}{\kappa})^2}
\end{equation}

Here, $\kappa$ is the inverse correlation length in the appropriate
direction. The data was fit using resolution
convolution techniques wherein the T=40 K peak served as the resolution function.  Inverse correlation lengths
extracted from this analysis along the longitudinal (open symbols) and
transverse (closed symbols) directions in reciprocal space are shown for
the (12,0,0) and (8,8,0) Bragg peaks of {\tto} in the middle and bottom panels of Fig. 3,
respectively.

Two features are clear from the measured inverse correlation lengths.  
First, the structure departs from long range order at temperatures as high
as 20 K, and spatial correlations diminish continuously as the temperature
is lowered to $\sim$ 0.3 K.  At the lowest temperatures, the spatial
correlations still comprise many ($\sim$ 80) unit cells.  Nevertheless,
the increase in peak width indicating the reduction in correlation length
is clear (see Fig. 2). Second, the broadening of the Bragg peaks is
anisotropic.  The longitudinal inverse correlation length at the (12, 0,
0) Bragg position and T=0.3 K is a factor of $\sim$ 2.5 greater than the 
transverse
inverse correlation length.  At the (8, 8, 0) Bragg position, the
anisotropy is reversed with a larger inverse correlation length along the
transverse direction at the lowest temperatures.

\begin{figure} 
\centering 
\includegraphics[width=8.5cm, height=8.5cm]{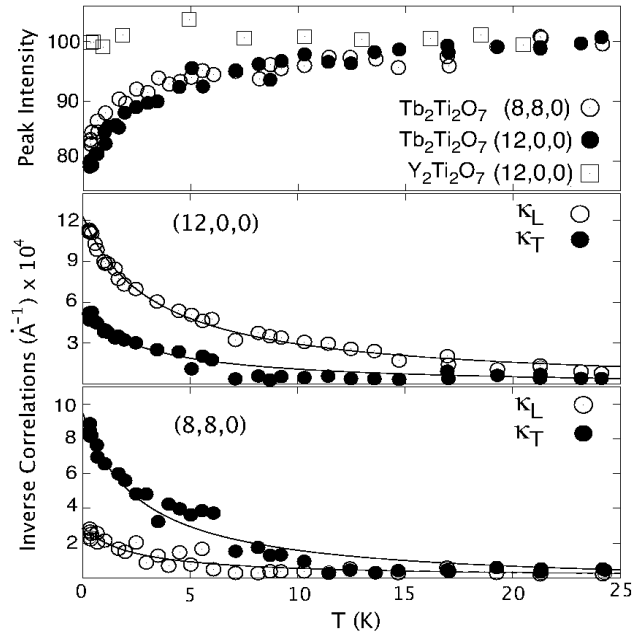} 
\caption{Top:  Temperature variaton of peak intensities in {\tto}
and {\yto} Middle:  Inverse correlation lengths in both principal
directions extracted from the (12, 0, 0) Bragg peak in {\tto}.  Bottom:  
Similarly extracted correlation lengths from the (8, 8, 0) Bragg peak in
{\tto}. The solid lines are guides to the eye.}
\label{Figure 3} 
\end{figure}

The anisotropy in the inverse correlation lengths at both the (12, 0, 0)
and (8, 8, 0) Bragg positions in {\tto} is qualitatively consistent with
cubic-to-tetragonal structural fluctuations preceding a cooperative
Jahn-Teller phase transition.  A non-zero temperature cubic-to-tetragonal
structural transition would introduce a second, distinct lattice parameter into
the structure.  This would result in the (12, 0, 0) reciprocal lattice
vector (with $\vert {\bf Q} \vert = 12 \times \frac{2\pi}{a}$) and the (0,
0, 12) reciprocal lattice vector (with $\vert {\bf Q} \vert =
12 \times \frac{2\pi}{c}$) splitting in the longitudinal direction in
reciprocal space.  Reciprocal lattice vectors of the form $\langle$8, 8, 0$\rangle$ would also show some splitting in the longitudinal direction, although it would be
smaller than that at the (12, 0, 0) Bragg position by a factor of $\sim$
3.  The principal splitting at the (8, 8, 0) Bragg position would be expected
in the transverse direction.

TbVO$_4$ and TbAsO$_4$ show related behavior near and below their
tetragonal-to-orthorhombic cooperative Jahn-Teller phase transitions at
33.1 K and 29.0 K, respectively\cite{Gehring, Hirano}.  Similar high
resolution x-ray scattering measurements performed with the same
experimental set up on these two materials show an extreme form of this
pattern of splitting of Bragg peaks below T$_C$ \cite{Lewis-Thesis}.  
These materials undergo the tetragonal-to-orthorhombic transition with 
twinning, such that the splitting of the
(12, 0, 0) Bragg peak is exclusively longitudinal, while that of the (8, 
8, 0) Bragg peak is exclusively transverse.  In addition, in a critical
temperature regime above T$_C$, Bragg peaks which ultimately split below
T$_C$ exhibit the same type of broadening (longitudinal for Bragg peaks of
the form (H, 0, 0) and transverse for Bragg peaks of the form (H, H, 0)).  
This is indicative of structural fluctuations above the cooperative
Jahn-Teller phase transition.  The anisotropic nature of both the
broadening (above T$_C$) and splitting (below T$_C$) of the Bragg peaks is 
more
extreme for TbVO$_4$ and TbAsO$_4$ than for the case of the correlation
lengths shown in Fig. 3 for {\tto}.  This is because the high temperature
phase in TbVO$_4$ and TbAsO$_4$ is tetragonal rather than cubic, and
measurements can focus on the unique (H,K,0) plane.

\begin{figure}
\centering
\includegraphics[width=8.5cm]{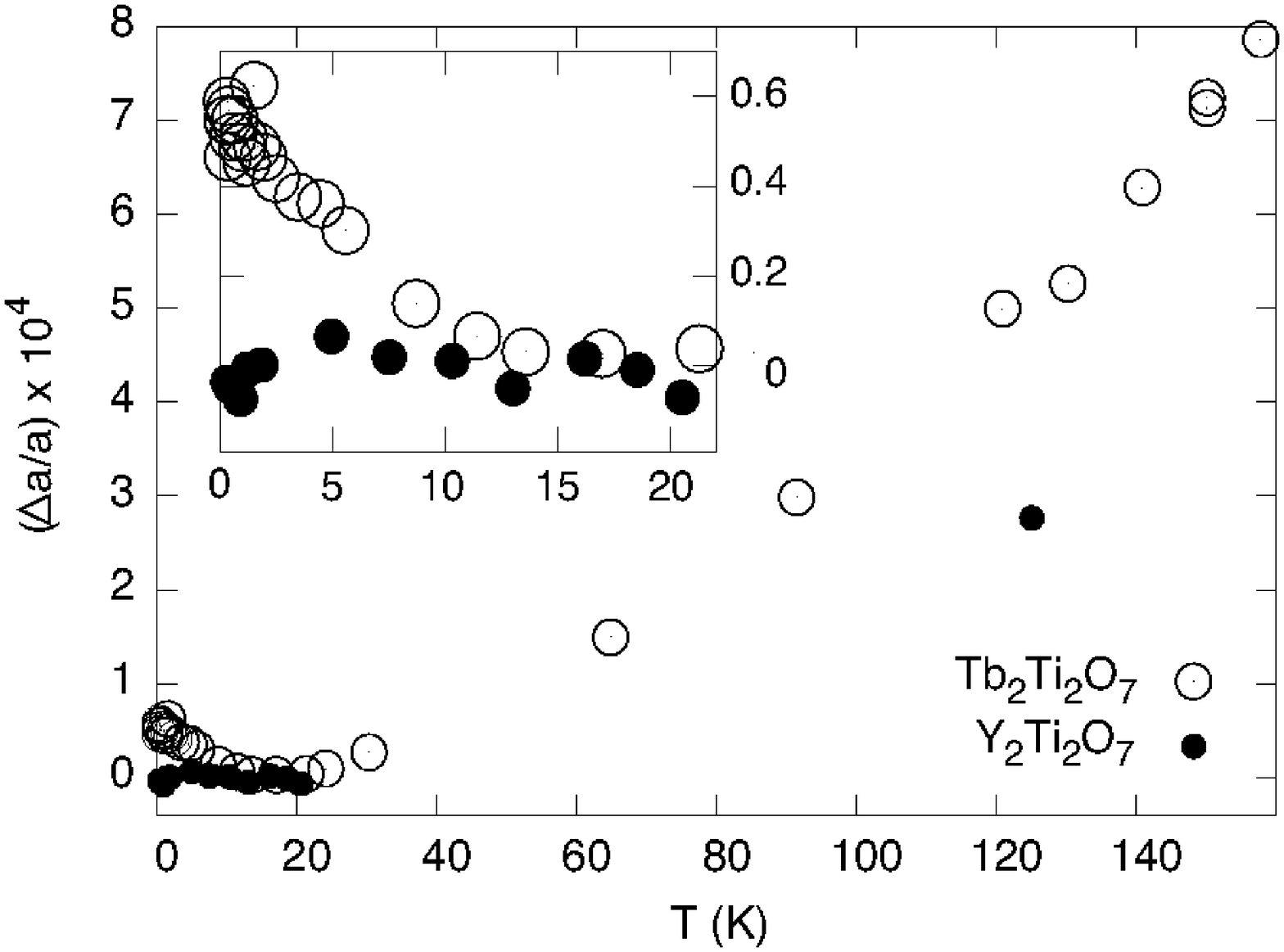}
\caption{Relative change in lattice constant as a function of temperature in both {\tto} and {\yto}.  Inset:  Focus on the anomalous
behaviour within the spin liquid regime of {\tto}.}
\label{Figure 4}
\end{figure}

Our high resolution x-ray measurements also allow an accurate
determination of the lattice parameter in {\tto} as a function of
temperature.  This is shown in Fig. 4, along with the corresponding
behavior of the lattice parameter in {\yto}.  From room
temperature to $\sim$ 18 K, we observe the normal thermal contraction of
the lattice in both {\tto} and {\yto}.  Interestingly, below $\sim$ 18 K,
the {\tto} lattice undergoes anomolous expansion which it maintains to our
base temperature of $\sim$ 0.3 K.  The inset to Fig. 4 highlights the
regime of thermal expansion with decreasing temperature at low
temperatures.  In contrast, the lattice parameter of non-magnetic {\yto}
is independent of temperature below 20 K, as is typical for most
materials.  While the unusual expansion of the lattice parameter at low
temperatures in {\tto} is small ($\frac{\Delta a}{a} < 10^{-4}$), it is both measureable and consistent
with earlier anomolous bulk lattice properties in the same low temperature
regime \cite{Mamsurova,Aleksandrov}, and coincident with the appearance of
the spin liquid or cooperative paramagnetic state.

An interesting question arising from this work is to what extent 
the fluctuating lattice symmetry could give rise to continuous spin 
correlations and dynamics, as opposed to local Ising correlations 
expected from the crystal field analysis.  We cannot draw definitive 
conclusions on this issue, but hope this work guides and informs
efforts to fully understand the enigmatic low temperature
magnetic properties of this frustrated antiferromagnet.

To conclude, high resolution x-ray scattering measurements on {\tto} show
continuous broadening below $\sim$ 20 K of allowed Bragg peaks, 
consistent with cubic-to-tetragonal fluctuations above a very low 
temperature cooperative
Jahn-Teller phase transition.  This fluctuation behaviour of the lattice
is coincident with the development of a highly correlated spin liquid or
cooperative paramagnetic ground state in {\tto}.  Unlike the case of the
cubic spinels, magnetoelastic coupling in {\tto} does not lead to a
satisfaction of spin interactions via a structural phase transition at
accessible temperatures.  Rather, the lowest temperature measurements
reveal a disordered ground state with coupled fluctuations in both the
spin and lattice degrees of freedom.  

We acknowledge useful conversations with G.M. Luke and M.J.P. Gingras.  
This work was supported by NSERC of Canada.


%
%






\end{document}